\documentclass[twocolumn]{aastex62}

\usepackage{amsfonts,amsbsy}
\usepackage{mathrsfs}
\usepackage{bm} 

\usepackage{graphicx} 
\usepackage{dcolumn} 

\usepackage{aas_macros}

\usepackage{xcolor}
\usepackage{color}

\usepackage{url}
\usepackage{multirow}

\usepackage{ulem}
\usepackage{enumerate}
\usepackage{textcmds}
\usepackage{mathtools}
\usepackage{booktabs}
\usepackage{tabularx}

\renewcommand{\vec}[1]{\boldsymbol{#1}}

\newcommand{\dif}{\mathrm{d}}

\newcolumntype{p}{D{,}{\pm}{-1}}

\renewcommand{\figurename}{Fig.}
\renewcommand{\tablename}{Tab.}

\renewcommand{\arraystretch}{1.5}

\begin{document}

\title{Revisiting Spatial-Dependent Propagation Model with Latest Observations of Cosmic Ray Nuclei}

\correspondingauthor{Wei Liu, Yu-hua Yao, Yi-Qing Guo}
\email{liuwei@ihep.ac.cn, yaoyh@ihep.ac.cn, guoyq@ihep.ac.cn}

\author{Wei Liu}
\affiliation{Key Laboratory of Particle Astrophysics,
Institute of High Energy Physics, Chinese Academy of Sciences, Beijing 100049, China
}

\author{Yu-hua Yao}
\affiliation{College of Physical Science and Technology, Sichuan University, Chengdu, Sichuan 610064, China
}
\affiliation{Key Laboratory of Particle Astrophysics,
Institute of High Energy Physics, Chinese Academy of Sciences, Beijing 100049, China
}

\author{Yi-Qing Guo}
\affiliation{Key Laboratory of Particle Astrophysics,
Institute of High Energy Physics, Chinese Academy of Sciences, Beijing 100049, China
}


\begin{abstract}
Recently AMS-02 collaboration publish their measurements of light cosmic-ray nuclei, including lithium, beryllium, boron, carbon and oxygen. All of them reveal a prominent excess above $\sim200$ GV, coinciding with proton and helium. Particularly, the secondary cosmic rays even harden than the primary components above that break. One of the viable interpretations for above anomalies is the spatial-dependent diffusion process. Such model has been successfully applied to multiple observational phenomena, for example primary cosmic ray nuclei, diffuse gamma ray and anisotropy. In this work, we investigate the spatial-dependent propagation model in light of the new observational data. We find that such model is able to explain the upturn of secondary spectrum as well as the primary's. All the spectra can be well reproduced and the calculated ratios are also in good agreement with the observations.
\end{abstract}


\section{Introduction} 
\label{sec:intro}

Nowadays, signs are growing that the conventional cosmic ray (CR) picture needs modifications. Among them, the unexpected hardening of CR proton and helium spectra above a few hundred GeV/nucleon has received much attention. This anomaly is observed by ATIC-2 \citep{2006astro.ph.12377P, 2009BRASP..73..564P}, CREAM \citep{2010ApJ...714L..89A, 2011ApJ...728..122Y, 2017ApJ...839....5Y} and PAMELA \citep{2011Sci...332...69A} experiments. According to their measurements, above $\sim200$ GeV/nucleon, the fluxes of both elements steadily rise with increasing energies. Later on, such feature is identified by the AMS-02 experiment with a higher significance \citep{2015PhRvL.114q1103A, 2015PhRvL.115u1101A}. Not long ago AMS-02 collaboration claim that both carbon and oxygen spectra also exhibit the same upturn above $\sim200$ GV, alike to helium \citep{PhysRevLett.119.251101}.

The excess also goes for the secondary CRs. AMS-02 collaboration assert the antiproton-to-proton ratio does not show explicit rigidity dependence from $\sim60$ to $\sim500$ GV \citep{2016PhRvL.117i1103A}. In other words, the antiproton flux has the identical rigidity dependence to the proton within this energy range. However the standard transport scenario predicts that the ratio of secondary-to-primary approximately falls off with the energy, i.e. $\Phi_s/\Phi_p \propto E^{-\delta}$, where $\delta$ is the power index of diffusion coefficient. In particular, most recently AMS-02 collaboration publish their detections of light secondary CR nuclei, including lithium, beryllium, and boron with very high accuracy \citep{PhysRevLett.120.021101}. The deviation from a single power-law clearly visible in all three spectra above $200$ GV, with an average hardening of $0.13 \pm 0.03$, more than the primaries.


The finding of spectral hardening brings about various alternatives of the traditional CR theory. Most of them fall into, but not limited to, three categories: acceleration process \citep{2011ApJ...729L..13O, 2012PhRvL.108h1104M, 2010ApJ...725..184B, 2011PhRvD..84d3002Y, 2017ApJ...835..229K}, transport effect \citep{2012PhRvL.109f1101B, 2012ApJ...752L..13T, 2012JCAP...01..010B, 2012ApJ...752...68V, 2014A&A...567A..33T, 2014ApJ...782...36E, 2015A&A...583A..95A, 2015PhRvD..92h1301T, 2015arXiv150908227G, 2016PhRvD..94l3007F, 2016ApJ...819...54G, 2016ChPhC..40a5101J, 2017arXiv170107136G, 2018arXiv180105904G}, and nearby source \citep{2012A&A...544A..92B, 2012MNRAS.421.1209T, 2013MNRAS.435.2532T, 2013A&A...555A..48B, 2015RAA....15...15L, 2015ApJ...803L..15T, 2015ApJ...815L...1T, 2017PhRvD..96b3006L, PhysRevLett.120.041103}. One of the popular scenarios is so-called spatial-dependent propagation (SDP) model. It is first introduced by \cite{2012ApJ...752L..13T} as Two Halo model (THM) and well explains the proton and helium anomaly. Usually the diffusion is regarded as only a function of rigidity, which does not vary spatially. But in SDP model, the whole transport volume is divided into two regions. The Galactic disk and its surrounding area are called the inner halo (IH), in which the diffusion property is influenced by the distribution of CR sources. Outside of IH, the diffusion approaches to the traditional assumption, i.e. only rigidity dependent. This extensive region is named as outer halo (OH). To reproduce the high energy excess, the diffusion coefficient within IH has a weaker rigidity dependence on average, compared with OH zone. In addition to the spectral hardening, the SDP model is also applied to solve the puzzles of anisotropy, diffuse gamma ray and so on \citep{2012ApJ...752L..13T, 2016ApJ...819...54G, 2018arXiv180105904G}.


The SDP model could also induce the bending of secondaries as well as the primaries. Given the new observations of both primary and secondary CRs, in this work we revisit the SDP model. We find that both primary and secondary spectra can be well described, and the ratios of B/C and $\rm \bar{p}/p$ are consistent with the available observations. The other ratios, such as Li/C, Be/C, Li/O, Be/O, Li/B and Be/B, are computed as well, which also agree with the measurements.

The rest of paper is organized as follows: in section \ref{sec:model}, we touch on the SDP model briefly. The results and some discussions are presented in section \ref{sec:res}. Finally section \ref{sec:concl} is reserved for our conclusion.

\section{Spatial-Dependent Propagation}
\label{sec:model}


After escaping from the acceleration sites, CRs undergo diffusive motion within so-called magnetic halo. It is usually approximated to a cylinder, with the Galactic disk embedded in the middle horizontally. The radial boundary of magnetic halo is identical to the Galactic radius, whereas its half  thickness $z_h$ is uncertain, which is determined by the CR data. Both CR sources and interstellar medium (ISM) chiefly spread within the Galactic disk. Aside from diffusion effect, CR nuclei may also suffer from the convection, re-acceleration as well as the fragmentation due to the collisions with interstellar gas. At lower energy, CR nuclei still lose their energy via ionization and Coulomb scattering. The transport equation is in general written as
\begin{eqnarray}
\frac{\partial \psi}{\partial t} &=& Q(\vec{r}, p) + \nabla \cdot ( D_{xx}\nabla\psi - \vec{V}_{c}\psi )
+ \frac{\partial}{\partial p}\left[p^2D_{pp}\frac{\partial}{\partial p}\frac{\psi}{p^2}\right]
\nonumber\\
&& - \frac{\partial}{\partial p}\left[ \dot{p}\psi - \frac{p}{3}(\nabla\cdot\vec{V}_c)\psi \right]
- \frac{\psi}{\tau_f} - \frac{\psi}{\tau_r} ~,
\label{propagation_equation}
\end{eqnarray}
where $\psi = \dif n/\dif p$ is the CR density per total particle momentum $p$ at position $\vec{r}$. On the halo border, the free escape condition is imposed, namely $\psi(R, z, p) =  \psi(r, \pm z_h, p) = 0$. For a comprehensive introduction to the CR transport, one can refer to \cite{2002astro.ph.12111M, 2007ARNPS..57..285S, 2015ARA&A..53..199G}. In this work, we adopt the well-known diffusion-reacceleration model, which is shown to well describe the spectra of both CR nuclei and secondary-to-primary ratios \citep{2011ApJ...729..106T, 2016ApJ...824...16J, 2017PhRvD..95h3007Y}.

The spatial distribution of CR sources is described by 
\begin{equation}
f(r, z) = \left(\dfrac{r}{r_\odot} \right)^\alpha \exp \left[-\dfrac{\beta(r-r_\odot)}{r_\odot} \right] \exp \left(-\dfrac{|z|}{z_s} \right) ~,
\label{eq:source_dis}
\end{equation}
where $r_\odot = 8.5$ kpc, $z_s = 0.2$ kpc, $\alpha = 1.09$, and $\beta = 3.87$ \citep{1998ApJ...504..761C} respectively. At the location of solar system, $f(r, z)$ is normalized to $1$. The rigidity dependence of the source term $Q(r, p)$ is assumed to be a broken power-law, i.e.
\begin{equation}
  q(\mathcal R) =  q_0 \times\left\{ \begin{array}{ll}
    \left( \dfrac{\mathcal R}{\mathcal R_{\rm br}} \right)^{-\nu_1} ~, & \mathcal R \leqslant \mathcal R_{\rm br}\\
    \left( \dfrac{\mathcal R}{\mathcal R_{\rm br}} \right)^{-\nu_2} \exp\left[-\dfrac{\mathcal R}{\mathcal R_{\rm c}} \right] ~, & \mathcal R > \mathcal R_{\rm br}
  \end{array}
  \right. ~,
\label{inject_spec_nuclei}
\end{equation}
where $q_0$ is the normalization factor. $\mathcal R_{\rm br}$ and $\mathcal R_{\rm c}$ are the broken and cutoff rigidity respectively. $\nu_1$($\nu_2$) is the spectral index below(above) $\mathcal R_{\rm br}$. The elemental abundance of CR sources is assigned according to the default value in the DRAGON package. 


Nearby the Galactic disk, the level of turbulence is appreciably affected by the activities of supernova explosions. Hence the turbulence is anticipated to be intense near the large population of sources so that the diffusion process is slow. At region with fewer sources, the turbulence is moderate and the diffusion tends to be fast. Therefore inside the IH, $D_{xx}$ is parameterized as
\begin{equation}
D_{xx}(r, z, \mathcal R) = F(r, z) D_0 \beta  \left(\dfrac{\mathcal R}{\mathcal R_0} \right)^{F(r, z) \delta_0} ~.
\end{equation}
$F(r,z)$ is anti-correlated with the source density, and its form is taken from \cite{2018arXiv180105904G}. The size of IH region is represented by its half thickness $\xi z_h$, whereas the OH region's is $(1- \xi)z_h$. Within the OH region, the turbulence is regarded as CR-driven in principle, which is less impacted by stellar activities.  Thus it is only rigidity dependent, $D_{xx} = D_0 \beta (\mathcal R/ \mathcal R_0)^{\delta_0}$.



Secondary particles, such as lithium, beryllium, boron and antiproton, are brought forth throughout the transport by the spallation and radioactive decay.  For the production of lithium, beryllium and boron, the so-called straight ahead approximation is adequate, in which the kinetic energy per nucleon is conserved during the interaction. Their source terms read
\begin{equation}
Q_{j = \rm Li, Be, B} = \sum_{i} (n_{\rm H} \sigma_{i+{\rm H}\rightarrow j} +n_{\rm He} \sigma_{i+{\rm He} \rightarrow j} ) v \psi_i ~,
\end{equation}
where $n_{\rm H/He}$ is the number density of interstellar hydrogen/helium and $\sigma_{i+{\rm H/He} \rightarrow j}$ is the total cross section of the corresponding hadronic interaction. Unlike above light nuclei, the generation of antiproton is  expressed as a convolution of the primary spectra $\psi_i(p)$ with the relevant differential cross section $d \sigma_{i + {\rm H/He} \to j}/d E_j$, i.e.
 \begin{eqnarray}
\nonumber Q_j &=& \sum_{i = \rm p, He} \int dp_i v \left\lbrace n_{\rm H} \frac{\dif \sigma_{i+{\rm H}\rightarrow j}(p_i, p_j)}{\dif p_j} \right. \\
& & \left. +n_{\rm He} \frac{\dif \sigma_{i+{\rm He}\rightarrow j}(p_i, p_j)}{\dif p_j} \right\rbrace  \psi_i(p_i) ~.
\end{eqnarray}

The numerical package DRAGON\footnote{https://github.com/cosmicrays} \citep{1475-7516-2008-10-018, 2017JCAP...02..015E} is introduced to solve the transport equation to obtain the spatial distribution of both primary and secondary CRs. Below tens of GeV, the CR flux are impacted by the solar modulation, so the well known force-field approximation is fetched to describe its effect, with modulation potential $\phi$ adjusted to fit low energy data.

\section{Results}
\label{sec:res}

\subsection{Primaries}
\label{subsec:pri}

\begin{figure*}
\centering
\includegraphics[height=14.cm, angle=0]{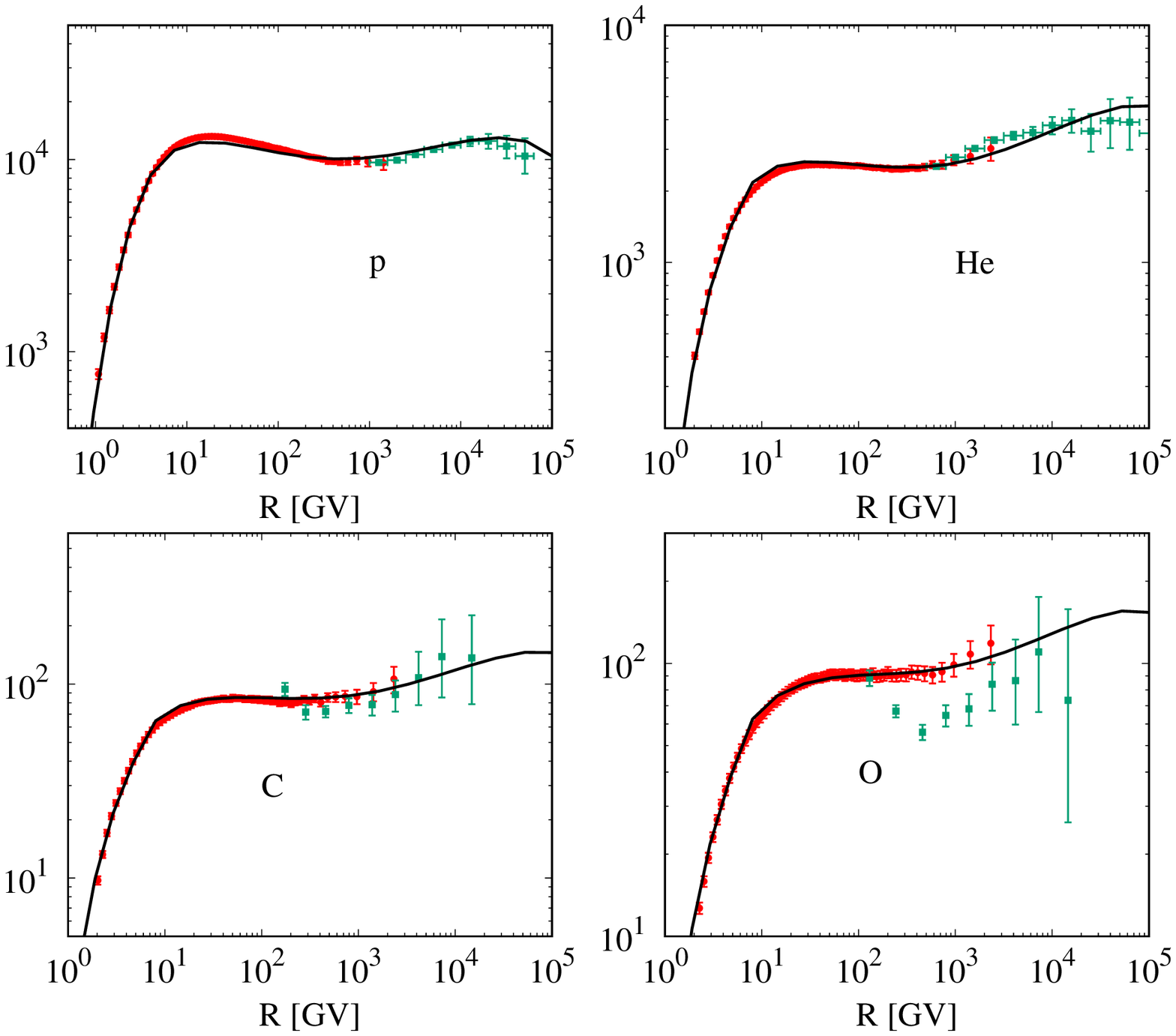}
\caption{
The primary CR energy spectra, i.e. proton (upper left), helium (upper right), carbon (lower left) and oxygen (lower right), in which each spectrum is multiplied by $\mathcal R^{2.7}$. Red data points denote the AMS-02 measurements \citep{2015PhRvL.114q1103A, PhysRevLett.119.251101}, while the green squares are from CREAM experiment \citep{2009ApJ...707..593A, 2017ApJ...839....5Y}. The black solid line are the calculated fluxes by SDP model. The parameters for transport and injection spectrum are listed in table \ref{tab:para_all}.
}
\label{fig:spec_pri}
\end{figure*}

 
Fig. \ref{fig:spec_pri} demonstrates the spectra of primary components as function of rigidity, including proton, helium, carbon and oxygen. The black solid line is the calculated flux of each element by SDP model. The fitted parameters for transport and injection spectrum are listed in Tab. \ref{tab:para_all}, where $N_m$ and $n$ are the parameters in $F(r,z)$ \citep{2018arXiv180105904G}. The propagation of higher energy CRs is dominated by IH region, in which the rigidity dependence of diffusion coefficient is flatter than that in OH region. It can be seen that both proton and helium spectra grow hardening above $\sim200$ GV, which is in consistent with AMS-02 detection. The position of the spectral break and its intensity principally relies on the parameter $N_m$, as well as the size of IH region, i.e. parameter $\xi$. To fit the spectra below tens of GV, the modulation potential for proton and helium are $\phi_{\rm p} = 700$ and $\phi_{\rm He} = 600$ MV. In addition, according to the observation of CREAM \citep{2017ApJ...839....5Y}, both proton and helium spectra soften at $\sim20$ TeV. To account for such spectral change, $\mathcal R_{\rm c}$ is adjusted to $180$ TV.

Since the transport processes of primary nuclei are similar, the spectral hardening is expected to occur in all the primary components. In addition to proton and helium, the calculated fluxes of both carbon and oxygen indeed show remarkable excesses. Both of them well reproduce the observed hardening by the latest measurements from AMS-02 experiment. Both carbon and oxygen fluxes have the same rigidity of upturn as helium. To fit the low energy data, the modulation potentials for carbon and oxygen are $\phi_{\rm C} = 500$ and $\phi_{\rm O} = 400$ MV respectively.

\begin{table}[h]
\begin{center}
\caption{Parameters of transport and injection spectrum.}
\begin{tabular}{ccccc}
\hline\hline
   $D_0$ (cm$^2$ s$^{-1}$) & $5.6\times 10^{28}$ \\
   $\delta_0$              & 0.6                \\
   $v_A$ (km s$^{-1}$)     & 6                 \\
   $z_h$ (kpc)             & 5                 \\
   $N_m$                   & 0.24                \\
   $\xi$                   & 0.1               \\
   $n$                     & 4                   \\
   \hline
   $q_0$ $[({\rm m}^2\cdot {\rm sr}\cdot {\rm s}\cdot {\rm GeV})^{-1}]$ & $4.22\times 10^{-2}$ \\
   $\nu_1$ & $1.8$ \\
   $\nu_2$ & 2.4 \\
   $\mathcal R_{\rm br}$ [GV]  & $4.2$ \\
   $\mathcal R_{c}$ [TV]  & $180$ \\
\hline \hline
\end{tabular}
\label{tab:para_all}
\end{center}
\end{table}

\subsection{Secondaries}
\label{subsec:sec}

\begin{figure*}
\centering
\includegraphics[height=14.cm, angle=0]{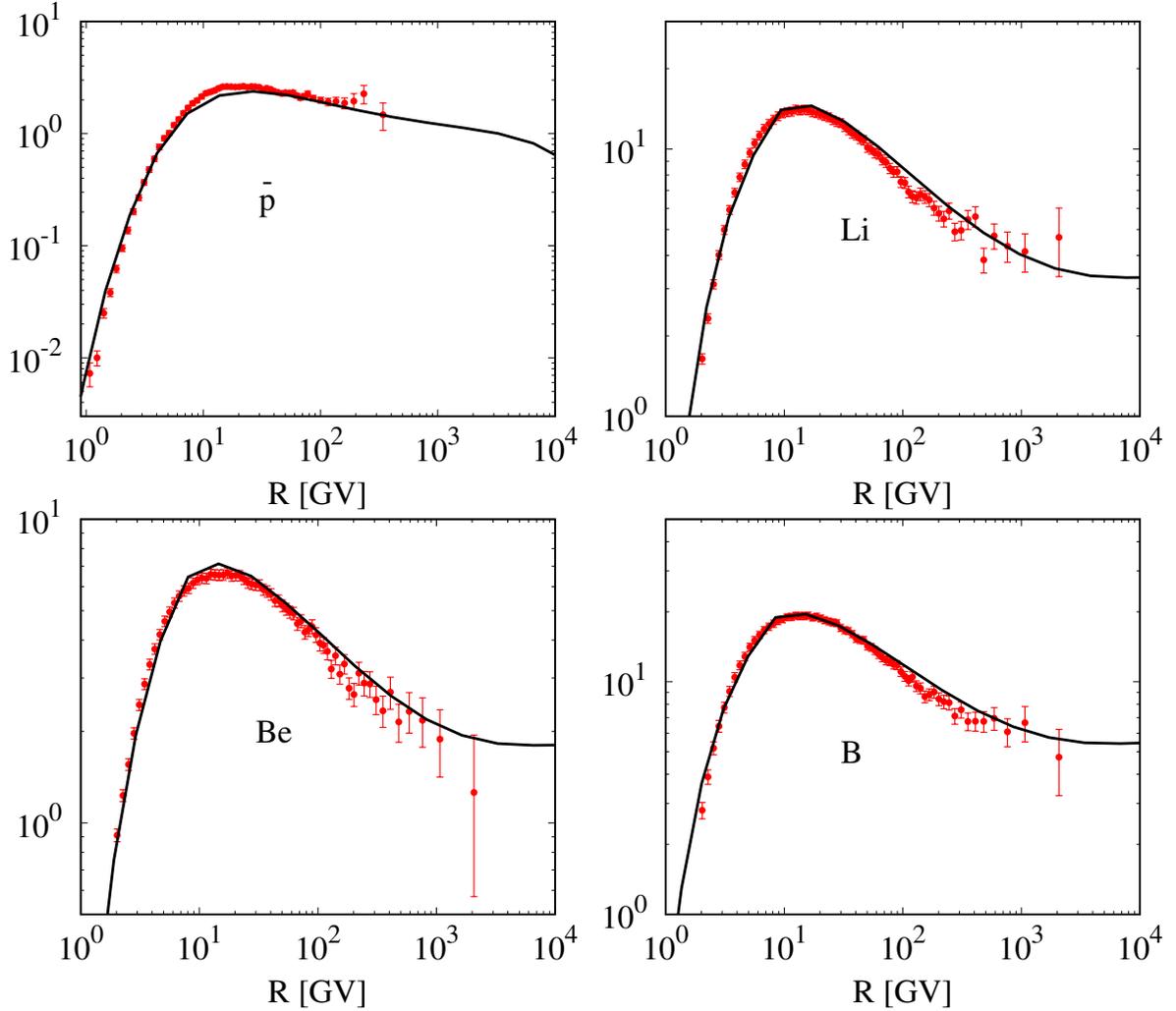}
\caption{
The secondary CR spectra, including antiproton (upper left), lithium (upper right), beryllium (lower left) and boron (lower right), in which each spectrum is multiplied by $\mathcal R^{2.7}$. Red data points denote the AMS-02 measurements \citep{2016PhRvL.117i1103A,  PhysRevLett.119.251101}.
}
\label{fig:spec_sec}
\end{figure*}

Similar excesses appear in the secondary CRs. In Fig. \ref{fig:spec_sec}, we show the comparison of the spectra computed by SDP model and the measurements as to the secondary CR nuclei. Since they are principally generated by light primary nuclei during the transport, the spectra of secondaries are expected to have the excess as well. Thanks to the straight-ahead approximation, the hardening of lithium, beryllium and boron are similar to their progenitors. As can be seen from the figure, all three spectra deviate from a single power law above $\sim200$ GV in the similar way, which is in accordance with the AMS-02 observation. Here the modulation potential for lithium and boron are $\phi_{\rm Li} = 550$ and $\phi_{\rm B} = 500$ MV, whereas for beryllium it is significantly larger, which needs $\phi_{\rm Be} = 900$ MV.

In contrast to lithium, beryllium and boron, the generated antiprotons by $\rm p-p$ and $\rm p-He$ interactions have a broad energy distribution.  In Fig. \ref{fig:spec_sec}, we could see that there is a slight rigidity dependence of antiproton flux above $10$ GV, which decreases gradually with energy. Within the uncertainty of measurements, our computation is compatible with the spectrum of antiproton. The modulation potential for antiproton is $\phi_{\rm \bar{p}} = 800$ MeV, which is a little different from proton.

\begin{figure*}
\centering
\includegraphics[height=7.cm, angle=0]{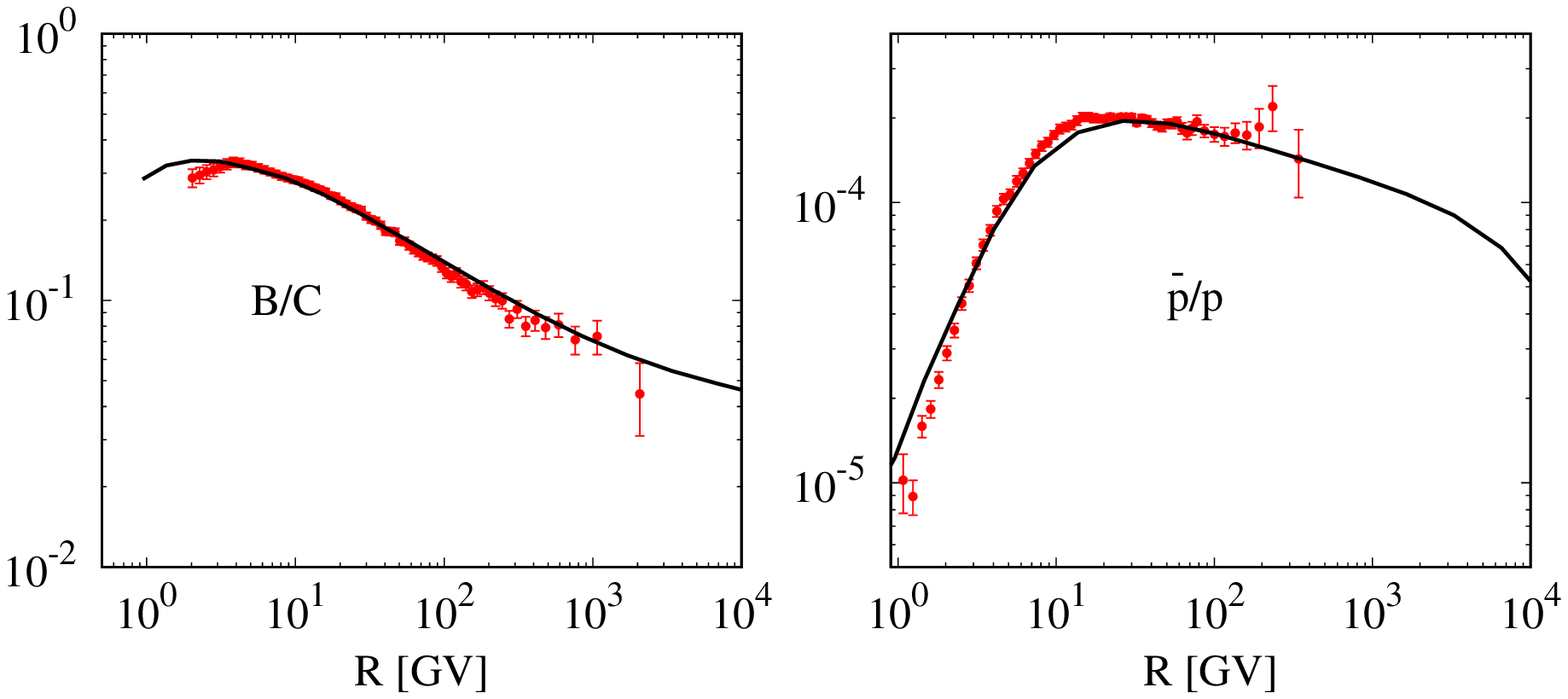}
\caption{
The B/C (left) and $\rm \bar{p}/p$ (right) ratios as function of rigidity. Red data points denote the AMS-02 measurements \citep{2015PhRvL.114q1103A, PhysRevLett.119.251101}.
}
\label{fig:spec_ratio}
\end{figure*}

We further compute the rigidity dependence of secondary-to-primary ratios, i.e. B/C and $\rm \bar{p}/p$, which is illustrated in Fig. \ref{fig:spec_ratio}. The B/C ratio is widely used to fix the transport parameters. Under the available parameters, SDP model can well describe the rigidity dependence of B/C ratio observed by AMS-02 in wide rigidity range, except below $3$ GV. Different from acceleration scenario, SDP model predicts that there is an excess of B/C ratio at higher energy, which is indicated in Fig. \ref{fig:spec_ratio}. Above $10^3$ GV, B/C ratio slowly becomes hardening. This can be testified by the accurate measurements above $1$ TV in the future. As for the $\rm \bar{p}/p$ ratio, although the spectrum calculated by SDP model can explain the observation within errors, there is still a slightly rigidity dependence. However according to the available measurements, this trend could not be distinguished from the pure flat spectrum. We hope future observation could give definite conclusion.

\subsection{Other Ratios}
\label{subsec:orat}

\begin{figure*}
\centering
\includegraphics[height=14.cm, angle=0]{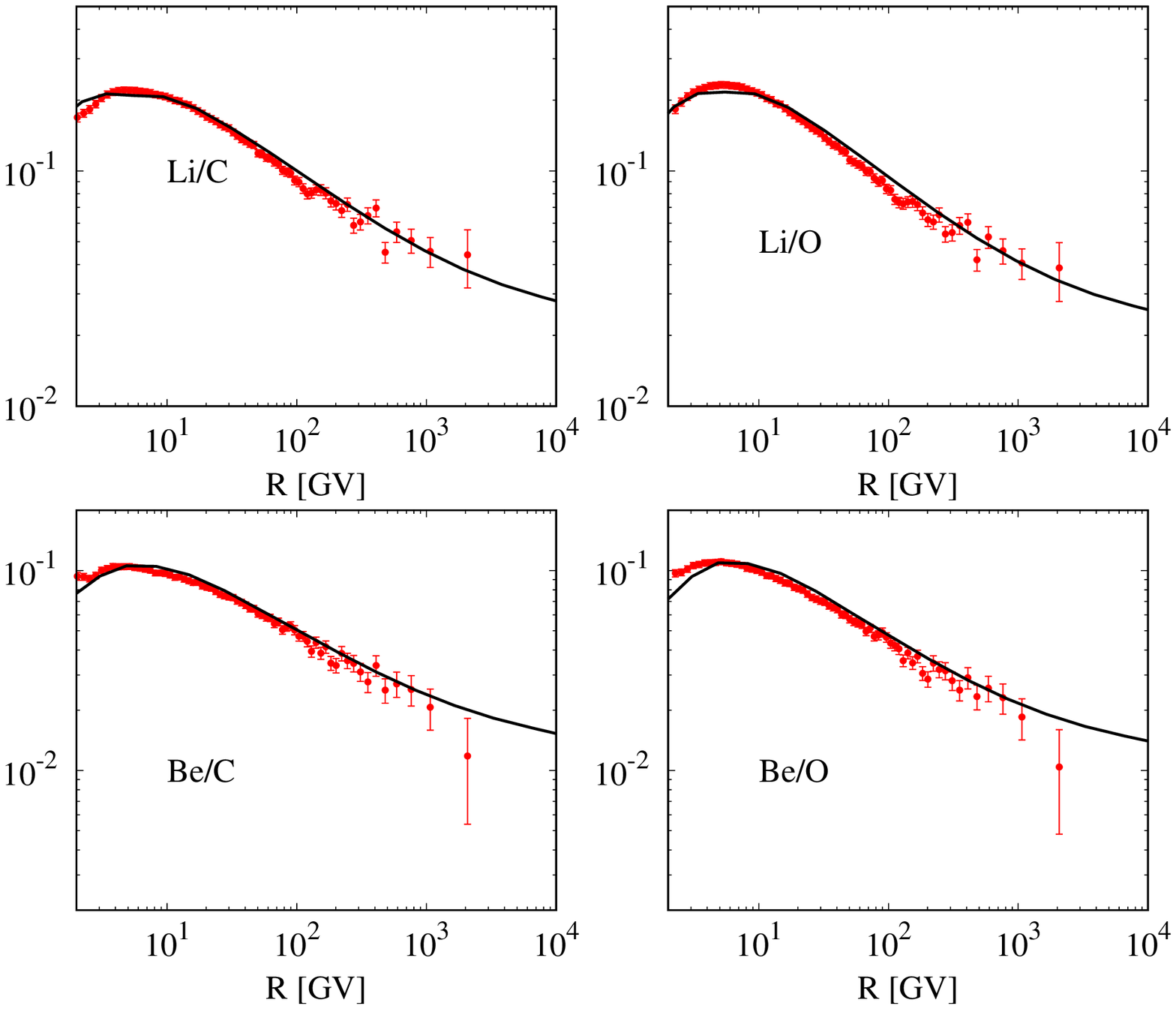}
\caption{
The ratios of secondary-to-primary CRs as function of rigidity, i.e. Li/C (upper left), Li/O (upper right), Be/C (lower left) and Be/O (lower right). Red data points denote the AMS-02 measurements \citep{PhysRevLett.119.251101}.
}
\label{fig:spec_ratio2}
\end{figure*}
\begin{figure*}
\centering
\includegraphics[height=7.cm, angle=0]{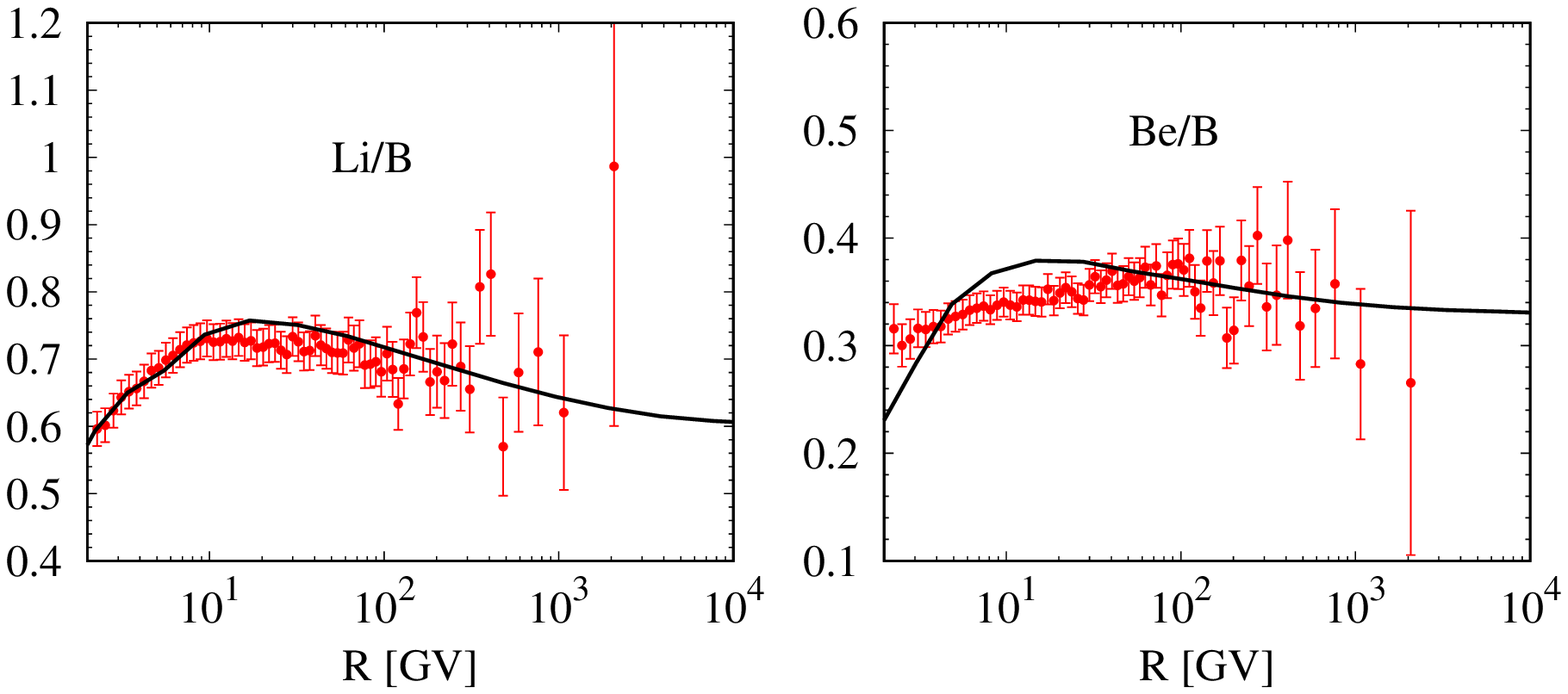}
\caption{
The ratios of secondaries as function of rigidity,  Li/B (left), Be/B (right). Red data points denote the AMS-02 measurements \citep{PhysRevLett.119.251101}.
}
\label{fig:spec_ratio3}
\end{figure*}

AMS-02 collaboration also release the spectra of other ratios. Fig. \ref{fig:spec_ratio2} illustrates the ratios of secondary-to-primary, i.e. Li/C, Be/C, B/O, Li/O, and Be/O respectively. Above $10$ GV, the rigidity dependence of each ratio is in consistent with AMS-02. Like B/C ratio, SDP model predicts there are excesses of these ratios above $1$ TV. We also compute the ratios of secondaries, e.g. Li/B and Be/B, as shown in Fig. \ref{fig:spec_ratio3}. The calculated Li/B ratio well reproduces the AMS-02 result. For Be/B,  below tens of GV, our computation has a harder rigidity dependence compared with measurements. Above $30$ GV, the rigidity dependence agrees with the AMS-02 data.

\section{Conclusion} 
\label{sec:concl}


With the development of instruments, the CR observations have been greatly improved. Recently AMS-02 collaboration successively issue their high-precision measurements of light CR nuclei, including lithium, beryllium, boron, carbon and oxygen. Along with the previously released proton and helium, the spectral hardening above $\sim200$ GV is widespread in both primaries and secondaries. These discoveries make people question the standard picture of CR acceleration and propagation.

In this work, we study the SDP model by the latest results from AMS-02. In SDP model, due to the smaller rigidity dependence in the vicinity of Galactic disk, the spectrum at higher energy becomes harder. We find that both primary and secondary spectra can be naturally reproduced and the B/C and $\rm \bar{p}/p$ ratios are in accordance with the measurements. Accurate observations could enable us to advance the understanding of CR acceleration and subsequent transport. We expect more accurate data can be available in the future to test SDP model. 


\acknowledgments

We thanks Prof. Hong-bo Hu and Qiang Yuan for helpful discussions. This work is supported by the National Key Research and Development Program of China (No. 2016YFA0400200), the National Natural Science Foundation of China (Nos. 11635011, 11761141001, 11663006).

\bibliographystyle{aasjournal}
\bibliography{ref}

\end{document}